\documentclass[floatfix,aps,amsmath,nofootinbib,onecolumn,11pt]{revtex4}

\pdfoutput=1
\usepackage{graphicx}
\usepackage{epstopdf}
\usepackage{amsmath}
\usepackage{amsfonts}
\usepackage{amssymb}
\usepackage{color}
\usepackage{mathrsfs}
\usepackage{multirow}
\usepackage{bm}

\def\({\left(}
\def\){\right)}
\def\[{\left[}
\def\]{\right]}

\def\e{\begin{equation}}
\def\q{\end{equation}}
\def\m{\begin{eqnarray}}
\def\n{\end{eqnarray}}

\begin{document}

\title{Constraints on the neutrino mass and mass hierarchy from cosmological observations}

\author{Qing-Guo Huang \footnote{huangqg@itp.ac.cn}, Ke Wang \footnote{wangke@itp.ac.cn} and Sai Wang \footnote{wangsai@itp.ac.cn}}
\affiliation{State Key Laboratory of Theoretical Physics, Institute of Theoretical Physics,
Chinese Academy of Science, Beijing 100190, People's Republic of China}

\date{\today}

\begin{abstract}

Considering the mass splitting between three active neutrinos, we represent the new constraints on the sum of neutrino mass $\sum m_\nu$ by updating the anisotropic analysis of Baryon Acoustic Oscillation (BAO) scale in the CMASS and LOWZ galaxy samples from Data Release 12 of the SDSS-III Baryon Oscillation Spectroscopic Survey (BOSS DR12). Combining the BAO data of 6dFGS, MGS, LOWZ and CMASS with $\textit{Planck}$~2015 data of temperature anisotropy and polarizations of Cosmic Microwave Background (CMB), we find that the $95\%$ C.L. upper bounds on $\sum m_\nu$ refer to $\sum m_{\nu,\rm NH}<0.18$ eV for normal hierarchy (NH), $\sum m_{\nu,\rm IH}<0.20$ eV for inverted hierarchy (IH) and $\sum m_{\nu,\rm DH}<0.15$ eV for degenerate hierarchy (DH) respectively, and the normal hierarchy is slightly preferred than the inverted one ($\Delta \chi^2\equiv \chi^2_{\rm NH}-\chi^2_{\rm IH} \simeq -3.4$). In addition, the additional relativistic degrees of freedom and massive sterile neutrinos are neither favored at present.

\end{abstract}

\pacs{????}

\maketitle

%%%%%%%%%%%%%%%%%%%%%%%%%%%%%%%%%%%%%%%%
%%%%%%%%%%%%%%%%%%%%%%%%%%%%%%%%%%%%%%%
\newpage
%%%%%%%%%%%%%%%%%%%%%%%%%%%%%%%%%%%%%%%
\section{Introduction}\label{introduction}
The phenomena of neutrino oscillation imply that there are mass splitting between three active neutrinos (see \cite{Lesgourgues:2006nd} for a review). Currently only two independent mass squared differences have been determined by neutrino oscillation experiments. Regardless of experimental uncertainties, they are given by \cite{pdg}
\begin{eqnarray}
\label{m21}
&&\Delta m_{21}^2\equiv m_2^2-m_1^2=7.5\times10^{-5}\textrm{eV}^2\ ,\\
\label{m31}
&&|\Delta m_{31}^2|\equiv|m_3^2-m_1^2|=2.5\times10^{-3}\textrm{eV}^2\ .
\end{eqnarray}
Thus we have two possible mass hierarchies, namely, a normal hierarchy (NH, $m_1<m_2<m_3$) and an inverted hierarchy (IH, $m_3<m_1<m_2$). Here $m_i$ ($i=1,2,3$) denote the mass eigenvalues of three neutrinos. The minimum sums of neutrino mass are $0.06$ eV for NH and $0.10$ eV for IH. Up to now, the absolute neutrino mass and mass hierarchy are still unknown.

Cosmology provides possibilities to measure the neutrino mass or the sum of neutrino mass $\sum m_\nu$ \cite{Hu:1997mj,Hu:2001bc,Komatsu:2008hk,Jimenez:2010ev,Reid:2009nq,Thomas:2009ae,Riemer-Sorensen:2013jsa,Swanson:2010sk,Cuesta:2015iho,Palanque-Delabrouille:2015pga,DiValentino:2015wba,DiValentino:2015sam,Rossi:2014nea,Hou:2012xq,Ade:2015xua,Gerbino:2015ixa,Zhang:2015rha,Zhang:2015uhk}. Massive neutrinos are initially relativistic and become non-relativistic today. They can impact on the cosmic expansion since they evolves differently from pure radiations and pure cold dark matter. They can influence the evolution of cosmological perturbations at early times and affect the CMB temperature anisotropies via the early-time Integrated Sachs-Wolfe (ISW) effect \cite{Hou:2012xq}. In addition, relativistic neutrinos suppress the clustering of matter and then modify the growth of structure. Thus one might extract useful signals of cosmic neutrinos from cosmological observations such as the matter clustering and the anisotropies and polarizations of Cosmic Microwave Background (CMB), etc.

$\textit{Planck}$ collaboration \cite{Ade:2015xua} gave the $95\%$ C.L. upper bounds on the total mass of three active neutrinos by assuming a degenerate hierarchy (DH, where $m_1=m_2=m_3$) regardless of the mass splitting. The $\textit{Planck}$ TT+lowP constraint is $\sum m_\nu<0.72$ eV and $\textit{Planck}$ TT,TE,EE+lowP constraint is $\sum m_\nu<0.49$ eV for the $\nu_{\rm DH}\Lambda$CDM model. Here TT denotes the power spectrum of CMB temperature, EE denotes the power spectrum of CMB E-mode, and TE denotes the CMB temperature and E-mode cross correlation in the $\textit{Planck}$~2015 data. ``lowP'' stands for $\textit{Planck}$ 2015 low-$\ell$ temperature-polarization data. Further adding the $\textit{Planck}$~2015 CMB lensing data \cite{Ade:2015zua}, the constraints are slightly changed to $\sum m_\nu<0.68$ eV and $\sum m_\nu<0.59$ eV for two data combinations, respectively. However, by contrast, adding the Baryon Acoustic Oscillation (BAO) data including 6dFGS \cite{Beutler:2011hx}, MGS \cite{Ross:2014qpa}, BOSS DR11 CMASS \cite{Anderson:2013zyy} and LOWZ \cite{Anderson:2013zyy} can significantly improve the constraints to $\sum m_\nu<0.21$ eV and $\sum m_\nu<0.17$ eV, respectively. The reason is that the BAO data can significantly break the acoustic scale degeneracy.

Recently the BAO distance scale measurements were updated via an anisotropic analysis of BAO scale in the correlation function \cite{Cuesta:2015mqa} and power spectrum \cite{Gil-Marin:2015nqa} of the CMASS and LOWZ galaxy samples from Data Release 12 of the SDSS-III Baryon Oscillation Spectroscopic Survey (BOSS DR12). The total volume probed in DR12 has a $10\%$ increment from DR11 and the experimental uncertainty has been reduced correspondingly. Thus in this paper we update the constraints on the total mass of three active neutrinos by using BOSS DR12 CMASS and LOWZ data, which are combined with other cosmological observations such as \textit{Planck}~2015 CMB data. In this paper, we also consider the mass splitting between three neutrinos implied by the neutrino oscillations between three generations. We estimate whether the current data sets can distinguish the neutrino mass hierarchy. In addition, we also update constraints on additional relativistic degree of freedom $\Delta N_{\textrm{eff}}\equiv N_{\textrm{eff}}-3.046$ and massive sterile neutrinos $m_{\nu,\textrm{sterile}}^{\textrm{eff}}$.

The rest of the paper is arranged as follows. In Sec.~\ref{dataset}, we reveal our methodology and cosmological data sets used in this paper. In Sec.~\ref{results}, we demonstrate our constraints on the sum of neutrino mass, additional relativistic degree of freedom and massive sterile neutrinos, respectively. Our conclusions are listed in Sec.~\ref{conclusion}.

\section{Data and Method}\label{dataset}
The recent distance measurements from the anisotropic analysis of BAO scale in the correlation function \cite{Cuesta:2015mqa} and power spectrum \cite{Gil-Marin:2015nqa} of CMASS and LOWZ galaxy samples from BOSS DR12 are listed in Tab.~\ref{BAO}.
\begin{table*}[!htp]
\centering
\renewcommand{\arraystretch}{1.5}
\begin{tabular}{ccccc}
\hline\hline
$z$     &BOSS DR12        &$H(z)r_{d}[10^3\textrm{km}\cdot s^{-1}]$     &$D_A(z)/r_{d}$    &$\rho_{D_A,H}$  \\
\hline
$0.32$                 &LOWZ        &$11.64\pm0.70$                            &$6.76\pm0.15$       &$0.35$    \\
\hline
$0.57$                 &CMASS       &$14.66\pm0.42$                            &$9.47\pm0.13$       &$0.54$    \\
\hline
\end{tabular}
\caption{The distance measurement from the anisotropic analysis of BAO scale in the CMASS and LOWZ galaxy samples released by SDSS-III BOSS DR12. Here we list the consensus values \cite{Gil-Marin:2015nqa}.}
\label{BAO}
\end{table*}
Only the consensus values \cite{Gil-Marin:2015nqa} are listed, which are used in this paper. Here $z$ denotes the effective redshift for CMASS and LOWZ samples, respectively, $H(z)$ and $D_A(z)$ are the Hubble parameter and angular diameter distance at reshift $z$ respectively, and $r_d$ is the comoving sound horizon at the redshift of baryon drag epoch. In addition, $\rho_{D_A,H}$ stands for the normalized correlation between $D_A(z)$ and $H(z)$.

In this paper, we combine the BAO data including 6dFGS \cite{Beutler:2011hx}, MGS \cite{Ross:2014qpa}, BOSS DR12 CMASS \cite{Gil-Marin:2015nqa} and LOWZ \cite{Gil-Marin:2015nqa} with $\textit{Planck}$~2015 likelihoods \cite{Aghanim:2015xee} of CMB temperature and polarizations as well as CMB lensing. In fact, we employ two combinations of observational data, namely $\textit{Planck}$ TT,TE,EE+lowP+BAO and $\textit{Planck}$ TT+lowP+lensing+BAO. The latter one is expected to give conservative constraints on the neutrino sectors while the former one gives more severe constraints. There are tensions on the amplitude of fluctuation spectrum between \emph{Planck} CMB data and other astrophysical data such as weak lensing (WL) \cite{Heymans:2012gg,Erben:2012zw}, redshift space distortion (RSD) \cite{Samushia:2013yga} and \emph{Planck} cluster counts \cite{Ade:2013lmv}. Thus we do not take them into consideration in this paper. We neither consider the direct measurements of cosmic expansion, since there are certain debates on the $H_0$ data \cite{Riess:2011yx,Freedman:2012ny,Efstathiou:2013via}. In addition, we do not use the data of supernovae of type Ia (SNe Ia), since the apparent magnitudes of SNe are insensitive to $\sum m_\nu$.

In the $\Lambda$CDM model, there are six base cosmological parameters which are denoted by \{$\omega_b$,$\omega_c$,$100\theta_{\textrm{MC}}$,$\tau,n_s$,$\textrm{ln}(10^{10}A_s)$\}. Here $\omega_b$ is the physical density of baryons today and $\omega_c$ is the physical density of cold dark matter today. $\theta_{\textrm{MC}}$ is the ratio between the sound horizon and the angular diameter distance at the decoupling epoch. $\tau$ is the Thomson scatter optical depth due to reionization. $n_s$ is the scalar spectrum index and $A_s$ is the amplitude of the power spectrum of primordial curvature perturbations at the pivot scale $k_p=0.05$ Mpc$^{-1}$.

To constrain the neutrino sectors, we refer to the Markov Chain Monte Carlo sampler (CosmoMC) \cite{Lewis:2002ah} in the $\nu\Lambda$CDM model.
By considering the mass splitting in Eq.~(\ref{m21}) and Eq.~(\ref{m31}), we can express the neutrino mass spectrum by two independent mass squared differences and one minimum mass eigenvalue $m_{\nu,{\textrm{min}}}$. The neutrino mass spectrum is
\begin{equation}(m_1,m_2,m_3)=(m_1,\sqrt{m_1^2+\Delta m_{21}^2},\sqrt{m_1^2+|\Delta m_{31}^2|})\end{equation} and $m_{\nu,{\textrm{min}}}=m_1$ for NH, and \begin{equation}(m_1,m_2,m_3)=(\sqrt{m_3^2+|\Delta m_{31}^2|},\sqrt{m_3^2+|\Delta m_{31}^2|+\Delta m_{21}^2},m_3)\end{equation} and $m_{\nu,{\textrm{min}}}=m_3$ for IH. In addition, the neutrino mass spectrum is trivial for DH, namely \begin{equation}m_1=m_2=m_3=m_{\nu,{\textrm{min}}}\ .\end{equation}
Thus we can constrain the sum of neutrino mass $\sum m_\nu$ via referring to the above three $\nu\Lambda$CDM model. It should be noted that there are lower cut-off values of $\sum m_\nu$ which are $0.06$ eV for NH and $0.10$ eV for IH, respectively.

\section{Results}\label{results}

In this section, we represent the constraints on the neutrino sectors by updating cosmological data. To be specific, we give an updated upper bound on the sum of neutrino mass $\sum m_\nu$ in Sec.~\ref{consumnu}. In Sec.~\ref{conneff}, the relativistic degree of freedom $N_{\textrm{eff}}$ is constrained. We simultaneously constrain $N_{\textrm{eff}}$ and massive sterile neutrino $m_{\nu,\textrm{sterile}}^{\textrm{eff}}$ in Sec.~\ref{conste}.

\subsection{Constraints on $\sum m_\nu$}\label{consumnu}

In this subsection, we refer to two combinations of data sets, namely $\textit{Planck}$ TT,TE,EE+lowP+BAO and $\textit{Planck}$ TT+lowP+lensing+BAO, to constrain the sum of neutrino mass $\sum m_\nu$ with NH, IH and DH, respectively. In the $\nu\Lambda$CDM model, the free cosmological parameters are given by
\begin{equation}
\{\omega_b,\omega_c,100\theta_{\textrm{MC}},\tau,n_s,\textrm{ln}(10^{10}A_s),m_{\nu,{\textrm{min}}}\}
\end{equation}
where $m_{\nu,{\textrm{min}}}$ is the minimal eigenvalue of neutrino mass, and the total mass of neutrinos is a derived parameter, i.e. $\sum m_\nu=m_1+m_2+m_3$.

\begin{figure}[]
\begin{center}
\includegraphics[width=8.0 cm]{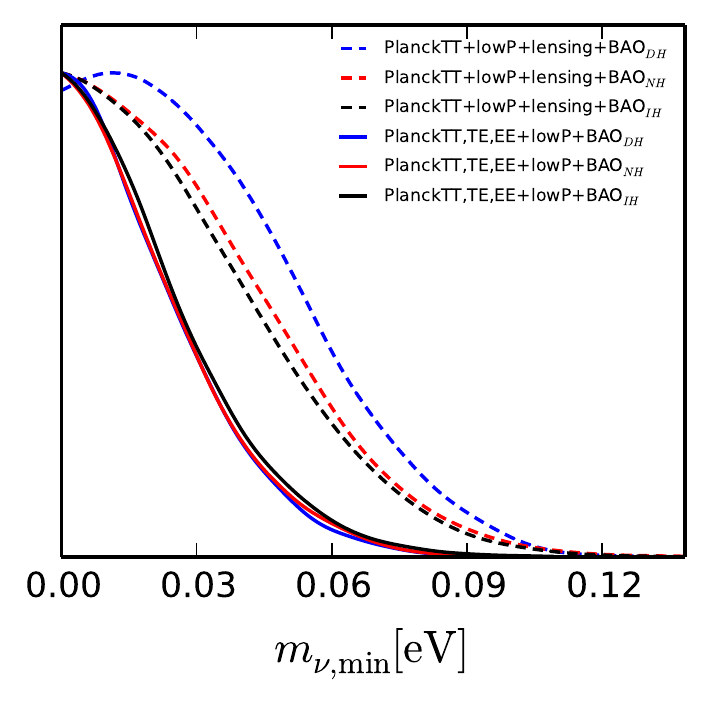}
\includegraphics[width=8.0 cm]{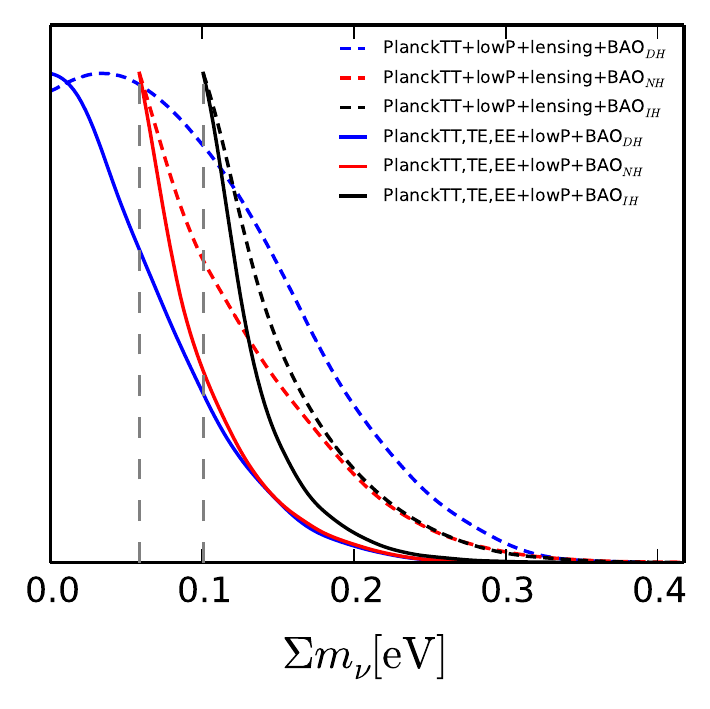}
\end{center}
\caption{The likelihood distributions of $m_{\nu,\textrm{min}}$ and $\sum m_\nu$ for the NH, IH and DH of neutrinos in the $\nu\Lambda$CDM models from two data combinations of $\textit{Planck}$ TT,TE,EE+lowP+BAO and $\textit{Planck}$ TT+lowP+lensing+BAO, respectively.}
\label{fig:sumnu}
\end{figure}

For three hierarchies, our constraints on $\sum m_{\nu}$ as well as seven free parameters and $\chi^2$ can be found in Tab.~\ref{tab:hierarchy}.
\begin{table*}[!htp]
\centering
\renewcommand{\arraystretch}{1.5}
\scalebox{0.8}[0.8]{%
\begin{tabular}{|c|c c c|c c c|}
\hline
&\multicolumn{3}{c|}{$\textit{Planck}$ TT,TE,EE+lowP+BAO}  &  \multicolumn{3}{c|}{$\textit{Planck}$TT+lowP+lensing+BAO}     \\
\cline{2-7}
&$\nu_{\textrm{NH}}\Lambda \textrm{CDM}$&$\nu_{\textrm{IH}}\Lambda \textrm{CDM}$&$\nu_{\textrm{DH}}\Lambda \textrm{CDM}$&$\nu_{\textrm{NH}}\Lambda \textrm{CDM}$&$\nu_{\textrm{IH}}\Lambda \textrm{CDM}$&$\nu_{\textrm{DH}}\Lambda \textrm{CDM}$\\
\hline
$\Omega_bh^2$&$0.02228\pm0.00014$&$0.02229\pm0.00014$&$0.02226\pm0.00014$&$0.02227\pm0.00020$ &$0.02229\pm0.00020$&$0.02226\pm0.00020$   \\
$\Omega_ch^2$&$0.1181\pm0.0011$&$0.1178\pm0.0011$&$0.1184\pm0.0011$&$0.1171\pm0.0013$&$0.1168\pm0.0013$&$0.1173\pm0.0013$ \\
$100\theta_{\emph{MC}}$&$1.04097\pm0.00030$&$1.04099\pm0.00030$&$1.04095\pm0.00030$&$1.04118\pm0.00040$ &$1.04120\pm0.00040$&$1.04118\pm0.00040$   \\
$\tau$&$0.085\pm0.017$&$0.088\pm0.017$&$0.082\pm0.017$&$0.076\pm0.015$&$0.080\pm0.015$&$0.072\pm0.016$\\
${\textrm{ln}}(10^{10}A_s)$&$3.101\pm0.033$&$3.105\pm0.033$&$3.095\pm0.033$&$3.079\pm0.029$&$3.086\pm0.028$&$3.073\pm0.031$  \\
$n_s$&$0.9657\pm0.0041$&$0.9664\pm0.0041$&$0.9650\pm0.0041$&$0.9684\pm0.0046$&$0.9690\pm0.0046$&$0.9677\pm0.0047$ \\
\hline
$m_{\textrm{min}}~(95\%)$&$<0.05~\textrm{eV}$&$<0.05~\textrm{eV}$&$<0.05~\textrm{eV}$&$<0.07~\textrm{eV}$&$<0.07~\textrm{eV}$&$<0.08~\textrm{eV}$   \\
$\sum m_\nu~(95\%)$&$<0.18~\textrm{eV}$&$<0.20~\textrm{eV}$&$<0.15~\textrm{eV}$&$<0.23~\textrm{eV}$  &$<0.25~\textrm{eV}$&$<0.23~\textrm{eV}$   \\
\hline
$\chi^2$&$12951.42$&$12954.80$&$12950.94$&$11283.67$&$11283.78$&$11284.34$ \\
\hline
\end{tabular}}
\caption{The $68\%$ limits for six base cosmological parameters and the $95\%$ limits for two neutrino mass parameters in the $\nu\Lambda$CDM models for the NH, IH and DH of neutrinos from two data combinations of $\textit{Planck}$ TT,TE,EE+lowP+BAO and $\textit{Planck}$ TT+lowP+lensing+BAO, respectively.}
\label{tab:hierarchy}
\end{table*}
The likelihood distributions of $\sum m_\nu$ and $m_{\nu,{\textrm{min}}}$ are depicted in Fig.~\ref{fig:sumnu}.
The dashed lines denote constraints from $\textit{Planck}$ TT+lowP+lensing+BAO dataset while the solid lines denote constraints from $\textit{Planck}$ TT,TE,EE+lowP+BAO dataset. The red, black and blue lines denote constraints for the NH, IH and DH of neutrino mass spectrum, respectively. The grey dashed lines denote the minimum values for the total mass of three neutrinos for NH and IH, respectively.

For the DH, the $95\%$ C.L. upper limit on the total mass of three active neutrinos is $\sum m_\nu<0.15$ eV for the data combination of $\textit{Planck}$ TT,TE,EE+lowP+BAO. The best-fit likelihoods is $\chi^2_{\rm DH}=12950.94$. Compared to $\textit{Planck}$~2015 constraint $\sum m_\nu<0.17$ eV in \cite{Ade:2015xua} from the $\textit{Planck}$ TT,TE,EE+lowP+BAO data where the BOSS DR11 CMASS and LOWZ data are used, there is about $10\%$ improvement on the uncertainty. The reason is that the total volume probed in BOSS DR12 has a $10\%$ increment and the experimental uncertainties are improved correspondingly. A conservative estimate is $\sum m_\nu<0.23$ eV with the best-fit likelihood $\chi^2_{\rm DH}=11284.34$ from the data combination of $\textit{Planck}$ TT+lowP+lensing+BAO.

For the NH, our constraint on $\sum m_\nu$ is given by $\sum m_\nu<0.18$ eV at $95\%$ C.L. from $\textit{Planck}$ TT,TE,EE+lowP+BAO dataset. It is around $20\%$ looser than the above constraint for the DH from the same dataset. The best-fit likelihoods for NH is $\chi^2_{\rm NH}=12951.42$ which is slightly larger than that for DH. On the other hand, our constraint becomes $\sum m_\nu<0.23$ eV at $95\%$ C.L. for $\textit{Planck}$ TT+lowP+lensing+BAO dataset, and the best-fit likelihood is $\chi^2_{\rm NH}=11283.67$. This constraint is similar to the constraint for DH by $\textit{Planck}$ TT+lowP+lensing+BAO dataset since this constraint is too loose to be sensitive to the neutrino mass hierarchy.

For the IH, our constraint on $\sum m_\nu$ is given by $\sum m_\nu<0.20$ eV at $95\%$ C.L. from $\textit{Planck}$ TT,TE,EE+lowP+BAO dataset. It is more than $30\%$ larger than that for the DH and about $10\%$ larger than that for the NH from the same dataset. The best-fit likelihoods is $\chi^2_{\rm IH}=12954.80$ which is larger than that for NH by $\Delta \chi^2=\chi^2_{\rm IH}-\chi^2_{\rm NH}\simeq 3.4$. It implies that the current data slightly prefers a normal hierarchy. On the other hand, our constraint becomes $\sum m_\nu<0.25$ eV at $95\%$ C.L. for $\textit{Planck}$ TT+lowP+lensing+BAO dataset.

\subsection{Constraints on $N_{\textrm{eff}}$}\label{conneff}

The total energy density of radiation in the Universe is given by
\begin{equation}
\rho=\[1+N_{\textrm{eff}}\frac{7}{8}\left(\frac{4}{11}\right)^{4/3}\]\rho_{\gamma}\ ,
\end{equation}
where $\rho_{\gamma}$ is the energy density of CMB photon and $N_{\textrm{eff}}=3.046$ for counting the standard model neutrinos. $N_{\textrm{eff}}>3.046$ will indicate that there are some unknown relativistic degrees of freedom in the Universe.

In this subsection, we use two data combinations of $\textit{Planck}$ TT,TE,EE+lowP+BAO and $\textit{Planck}$ TT+lowP+lensing+BAO to constrain $N_{\textrm{eff}}$ or equivalently the additional relativistic degree of freedom $\Delta N_{\textrm{eff}}=N_{\textrm{eff}}-3.046$ in the base $\Lambda$CDM+$N_{\textrm{eff}}$ model. The free parameters include six base parameters and $N_{\textrm{eff}}$, while we fix $\sum m_\nu=0.06$ eV with two massless and one massive active neutrinos. Our constraints on $N_{\textrm{eff}}$ can be found in Tab.~\ref{tab:neff}, where we also list constraints on other free parameters.
\begin{table*}[!htp]
\centering
\renewcommand{\arraystretch}{1.5}
\scalebox{0.8}[0.8]{%
\begin{tabular}{|c|c c|c c|}
\hline
&\multicolumn{2}{c|}{$\textit{Planck}$ TT,TE,EE+lowP+BAO}  &  \multicolumn{2}{c|}{$\textit{Planck}$ TT+lowP+lensing+BAO}     \\
\cline{2-5}
&$\Lambda$CDM+$N_{\textrm{eff}}$&$\Lambda$CDM+$N_{\textrm{eff}}$+$m_{\nu,\textrm{sterile}}^{\textrm{eff}}$&$\Lambda$CDM+$N_{\textrm{eff}}$&$\Lambda$CDM+$N_{\textrm{eff}}$+$m_{\nu,\textrm{sterile}}^{\textrm{eff}}$\\
\hline
$\Omega_bh^2$&$0.02230\pm0.00019$&$0.02241\pm0.00017$&$0.02230\pm0.00024$&$0.02249\pm0.00024$    \\
$\Omega_ch^2$&$0.1192\pm0.0031$&$0.1191\pm0.0033$&$0.1191\pm0.0037$&$0.1208\pm0.0038$    \\
$100\theta_{\emph{MC}}$&$1.04085\pm0.00044$&$1.04072\pm0.00034$&$1.04100\pm0.00056$&$1.04076\pm0.00051$    \\
$\tau$&$0.083\pm0.017$&$0.088\pm0.017$&$0.068\pm0.013$&$0.081\pm0.017$    \\
${\textrm{ln}}(10^{10}A_s)$&$3.099\pm0.035$&$3.112\pm0.035$&$3.068\pm0.026$&$3.100\pm0.035$    \\
$n_s$&$0.9667\pm0.0075$&$0.9700\pm0.0062$&$0.9698\pm0.0084$&$0.9781\pm0.0091$    \\
\hline
$N_{\textrm{eff}}~(95\%)$&$3.06\pm0.36$&$<3.39$&$3.10^{+0.45}_{-0.44}$&$<3.69$    \\
$m_{\nu,\textrm{sterile}}^{\textrm{eff}}~(95\%)$&--&$<0.60~\textrm{eV}$&--&$<0.48~\textrm{eV}$\\
\hline
\end{tabular}}
\caption{The $68\%$ limits for six base cosmological parameters and the $95\%$ limits for two neutrino parameters in the base $\Lambda$CDM+$N_{\textrm{eff}}$ and base $\Lambda$CDM+$N_{\textrm{eff}}$+$m_{\nu,\textrm{sterile}}^{\textrm{eff}}$ models from two data combinations of $\textit{Planck}$ TT,TE,EE+lowP+BAO and $\textit{Planck}$ TT+lowP+lensing+BAO, respectively.}
\label{tab:neff}
\end{table*}

Our results are well consistent with the standard prediction $N_{\textrm{eff}}=3.046$. The constraints on the effective number of relativistic degrees of freedom are $N_{\textrm{eff}}=3.06\pm0.36$ and $N_{\textrm{eff}}=3.10^{+0.45}_{-0.44}$ at $95\%$ C.L. from $\textit{Planck}$ TT,TE,EE+lowP+BAO and $\textit{Planck}$ TT+lowP+lensing+BAO, respectively. $\Delta N_{\textrm{eff}}=1$, for example a fully thermalized sterile neutrino, is excluded at more than $5\sigma$ level by $\textit{Planck}$ TT,TE,EE+lowP+BAO data and at $4\sigma$ level by $\textit{Planck}$ TT+lowP+lensing+BAO data.
A thermalized massless boson decoupled in the range $0.5$ MeV $<T<$ 100 MeV predicts $\Delta N_{\rm eff}=4/7\simeq 0.57$ which is disfavored at more than $95\%$ C.L. by these two data sets. If it decoupled at $T>$ 100 MeV, $\Delta N_{\rm eff}\simeq 0.39$ which is slightly disfavored by $\textit{Planck}$ TT,TE,EE+lowP+BAO data but slightly favored by $\textit{Planck}$ TT+lowP+lensing+BAO data.

%Our results are similar to $\textit{Planck}$~2015 results in \cite{Ade:2015xua}.

\subsection{Simultaneous constraints on $N_{\textrm{eff}}$ and $m_{\nu,\textrm{sterile}}^{\textrm{eff}}$}\label{conste}

We can also consider extra one massive sterile neutrino whose effective mass is parametrized by $m_{\nu,\textrm{sterile}}^{\textrm{eff}}\equiv\left(94.1\Omega_{\nu,\textrm{sterile}}h^2\right)$ eV. Assuming the sterile neutrino to be thermally distributed with an arbitrary temperature, $m_{\nu,\textrm{sterile}}^{\textrm{eff}}$ is then given by \begin{equation}m_{\nu,\textrm{sterile}}^{\textrm{eff}}=\left(\Delta N_{\textrm{eff}}\right)^{3/4}m_{\textrm{sterile}}^{\textrm{thermal}}\ ,\end{equation}
where $m_{\textrm{sterile}}^{\textrm{thermal}}$ denotes the true mass. Here we consider the base $\Lambda$CDM+$N_{\textrm{eff}}$+$m_{\nu,\textrm{sterile}}^{\textrm{eff}}$ model, in which $m_{\textrm{sterile}}^{\textrm{thermal}}$ is a free parameter with a prior $m_{\textrm{sterile}}^{\textrm{thermal}}<10$ eV and $N_{\textrm{eff}}$ has a flat prior with $N_{\textrm{eff}}>3.046$.

Our simultaneous constraints on $N_{\textrm{eff}}$ and $m_{\nu,\textrm{sterile}}^{\textrm{eff}}$ can be found in Tab.~\ref{tab:neff}. From $\textit{Planck}$ TT,TE,EE+lowP+BAO data, we obtain constraints to be $N_{\textrm{eff}}<3.39$ and ${m_{\nu,\textrm{sterile}}^{\textrm{eff}}}<0.60$ eV at $95\%$ C.L.. From $\textit{Planck}$ TT+lowP+lensing+BAO data, we obtain $N_{\textrm{eff}}<3.69$ and ${m_{\nu,\textrm{sterile}}^{\textrm{eff}}}<0.48$ eV at $95\%$ C.L., which are similar to $\textit{Planck}$~2015 results in \cite{Ade:2015xua}. $\Delta N_{\textrm{eff}}=1$ can be excluded at much more than $95\%$ C.L.. One should note that the upper tail of $m_{\nu,\textrm{sterile}}^{\textrm{eff}}$ is closely related to high physical masses near to the prior cutoff.

\section{Conclusions}\label{conclusion}

In this paper, we updated cosmological constraints on the total mass of three active neutrinos by updating BOSS DR11 to DR12 of CMASS and LOWZ samples. We considered the mass splitting between three neutrinos and then considered the neutrino mass spectrum with the NH, IH and DH, respectively. When the $\textit{Planck}$ TT,TE,EE+lowP+BAO combination is updated, our constraint $\sum m_\nu<0.15~\textrm{eV}$ at $95\%$ C.L. is improved by about $10\%$ for the DH, comparing to $\textit{Planck}$~2015 constraint $\sum m_\nu<0.17~\textrm{eV}$ at $95\%$ C.L. \cite{Ade:2015xua}. Meanwhile, we get updated $95\%$ C.L. upper limits $\sum m_\nu<0.18~\textrm{eV}$ for the NH and $\sum m_\nu<0.20~\textrm{eV}$ for the IH. For the NH (or the IH) and the DH, there is about $20\%$ (or $27\%$) difference between their upper limits on the absolute neutrino mass. Thus it is meaningful to take into consideration the data of neutrino mass splitting obtained from the experimental particle physics. Although the current cosmological data may be not good enough to distinguish different neutrino mass hierarchies, the normal hierarchy is slightly preferred by $\Delta\chi^2\simeq -3.4$ compared to the inverted hierarchy in our paper. Future precise observations might have potential to determine the neutrino mass and mass hierarchy \cite{Carbone:2010ik,Wong:2011ip,Hall:2012kg,Hamann:2012fe,Audren:2012vy,Font-Ribera:2013rwa,Abazajian:2013oma,Wu:2014hta,Mueller:2014dba,Zhen:2015yba,Villaescusa-Navarro:2015cca,Errard:2015cxa,Allison:2015qca,Liu:2015txa,Oyama:2015gma,Zhao:2015gua}.

There are various tight constraints on $\sum m_\nu$ in literatures. For instance, the combination of Lyman-$\alpha$ absorption in the distant quasar spectra, BAO and Planck CMB data gave a constraint $\sum m_\nu<0.12$ eV at $95\%$ C.L. in \cite{Palanque-Delabrouille:2015pga}. The combination of SDSS DR7 Luminous Red Galaxies (LRG), BAO and Planck CMB data gave an upper bound $\sum m_\nu<0.11$ eV at $95\%$ C.L. in \cite{Cuesta:2015iho}. Both constraints, close to the lower cut-off values of 0.10 eV for the IH, are tighter than ours obtained in this paper. Thus it is interesting to include the observational data sets regarding to the matter power spectrum into our exploration, besides the lensed-CMB and BAO data. We will remain these considerations as our future work.

In addition, we also updated the constraints on the relativistic degree of freedom and massive sterile neutrinos. Our results are similar to $\textit{Planck}$~2015 constraints in \cite{Ade:2015xua}. We found no significant evidence for additional relativistic degree of freedom and fully thermalized massive sterile neutrinos by using current data sets in this paper. Nevertheless, a significant density of additional radiations is still allowed by considering uncertainties of the data.

\vspace{0.5cm}
\noindent {\bf Acknowledgments}
We acknowledge the use of HPC Cluster of SKLTP/ITP-CAS. This work is supported by Top-Notch Young Talents Program of China and grants from NSFC (grant NO. 11322545, 11335012 and 11575271). QGH would also like to thank the participants of the advanced workshop ``Dark Energy and Fundamental Theory" supported by the Special Fund for Theoretical Physics from the National Natural Science Foundations of China (grant No. 11447613) for useful conversation.

%%%%%%%%%%%%%%%%%%%%%%%%%%%%%%%%%%%%%%%%
%%%%%%%%%%%%%%%%%%%%%%%%%%%%%%%%%%%%%%%%

%%%%%%%%%%%%%%%%%%%%%%%%%%%%%%%%%%%%%%%%
%%%%%%%%%%%%%%%%%%%%%%%%%%%%%%%%%%%%%%%%

%%%%%%%%%%%%%%%%%%%%%%%%%%%%%%%%%%%%%%%%
%%%%%%%%%%%%%%%%%%%%%%%%%%%%%%%%%%%%%%%%
\end{document}